\pdfoutput=1 
\documentclass[sigchi]{acmart}
\usepackage{threeparttable}
\usepackage{adjustbox}
\usepackage{booktabs} 
\setcopyright{none}
\acmDOI{}
\acmISBN{}

\acmConference[WSDM Cup 2019 Workshop]{ACM International Conference on Web Search and Data Mining}{February 2019}{Melbourne,  Austrailia}
\acmYear{2019}
\copyrightyear{2019}

\begin{document}
\title[]{Sequential Skip Prediction with Few-shot\\ in Streamed Music Contents}

\author{Sungkyun Chang}
\authornote{Sungkyun Chang is also with the Institute for Industrial Systems Innovation, Seoul National University, Seoul, Korea.}
\orcid{1234-5678-9012}
\affiliation{%
  \institution{Music and Audio Research Group\\ Center for Super Intelligence\\ Seoul National University}
  \streetaddress{P.O. Box 1212}
  \postcode{43017-6221}
}
\email{rayno1@snu.ac.kr}

\author{Seungjin Lee}
\orcid{1234-5678-9012}
\affiliation{%
  \institution{Music and Audio Research Group\\ Seoul National University}
  \streetaddress{P.O. Box 1212}
  \postcode{43017-6221}
}
\email{joshua77@snu.ac.kr}

\author{Kyogu Lee}

\orcid{1234-5678-9012}
\affiliation{%
  \institution{Music and Audio Research Group\\ Center for Super Intelligence\\ Seoul National University}
  \streetaddress{P.O. Box 1212}
  \postcode{43017-6221}
}
\email{kglee@snu.ac.kr}

\renewcommand{\shortauthors}{S. Chang et al.}

\begin{abstract}
This paper provides an outline of the algorithms submitted for the WSDM Cup 2019 Spotify Sequential Skip Prediction Challenge (team name: mimbres). In the challenge, complete information including acoustic features and user interaction logs for the first half of a listening session is provided. Our goal is to predict whether the individual tracks in the second half of the session will be skipped or not, only given acoustic features. We proposed two different kinds of algorithms that were based on metric learning and sequence learning. The experimental results showed that the sequence learning approach performed significantly better than the metric learning approach. Moreover, we conducted additional experiments to find that significant performance gain can be achieved using complete user log information.
\end{abstract}

%
%



\maketitle

\section{Introduction}
In online music streaming services such as {\it Spotify}\footnote{https://www.spotify.com/}, a huge number of active users are interacting with a library of over 40 million audio tracks. Here, an important challenge is to recommend the right music item to each user. To this end, there has  been a large related body of works in music recommender systems. A standard approach was to construct a global model based on user's play counts\cite{celma2010music, van2013deep} and acoustic features\cite{van2013deep}. However, a significant aspect missing in these works is how a particular user sequentially interacts with the streamed contents. This can be thought as a problem of  personalization\cite{cho2002personalized} with few-shot, or meta-learning\cite{snail} with external memory\cite{santoro2016meta}. The WSDM Cup 2019 tackles this issue by defining a new task with a real dataset\cite{brost2019music}. We can summarize the task as follows:
\begin{itemize}
\item The length $L^i$ of an $i$-th listening session for a blinded-particular user varies in the range from 10 to 20. We omit $i$ for readability from next page.
\item We denote the input sequence (Figure 1) from the first half (={\it support}) and second half(={\it query}) of each session $i$ as $X^i_s$ and  $X^i_q$, respectively.
\item $X^i_s$ contains complete information including session logs and acoustic features.
\item $X^i_q$ contains only acoustic features.
\item $Y^i_s$ is the labels representing whether the supports were skipped$(=1)$ or not$(=0)$.
\item Given a set of inputs $\{X^i_s, Y^i_s, X^i_q\}$, our task is to predict $Y^i_q$ (Figure 2).
\end{itemize}

One limitation of our research was that we did not make use of any external dataset nor pre-trained model from them. The code\footnote{https://github.com/mimbres/SeqSkip} and evaluation results\footnote{https://www.crowdai.org/challenges/spotify-sequential-skip-prediction-challenge} are available online.

\begin{figure*}
\includegraphics[width=6.1in]{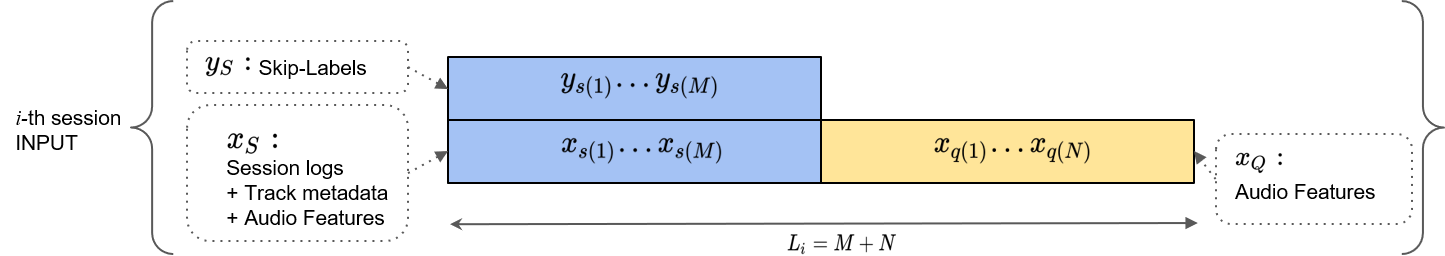}\Description{Several flies, spanning two
  columns of text}
\caption{Input structure; The blue and yellow blocks represent the inputs of supports and queries for prediction, respectively.}
\end{figure*}

\section{Model Architectures}
\begin{figure}
\includegraphics[width=3.1in]{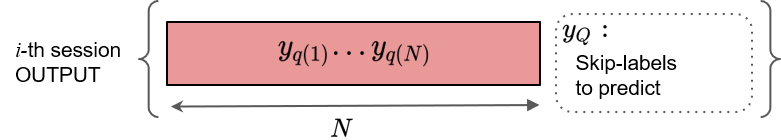}\Description{A fly image,
  to $1''\times1''$}
\caption{Output structure; The red block represents the skip-labels to be predicted for the $i$-th session.}
\end{figure}
In this section, we explain two different branches of algorithms based on 1) metric learning, and 2) sequence learning.
In metric learning-based approach, one key feature is that we do not assume the presence of orders in a sequence. This allows us to formulate the skip prediction problem in a similar way with the previous  works\cite{sung2018learning} on few-shot learning that {\it learns to compare}.

In sequence learning-based approach, we employ temporal convolution layers that can learn or memorize information by assuming the presence of orders in a sequence. In this fashion, we formulate the skip prediction problem as a meta-learning\cite{snail} that learns to {\it refer past experience}.  

\subsection{Metric Learning}

\begin{figure}
\includegraphics[width=3.2in]{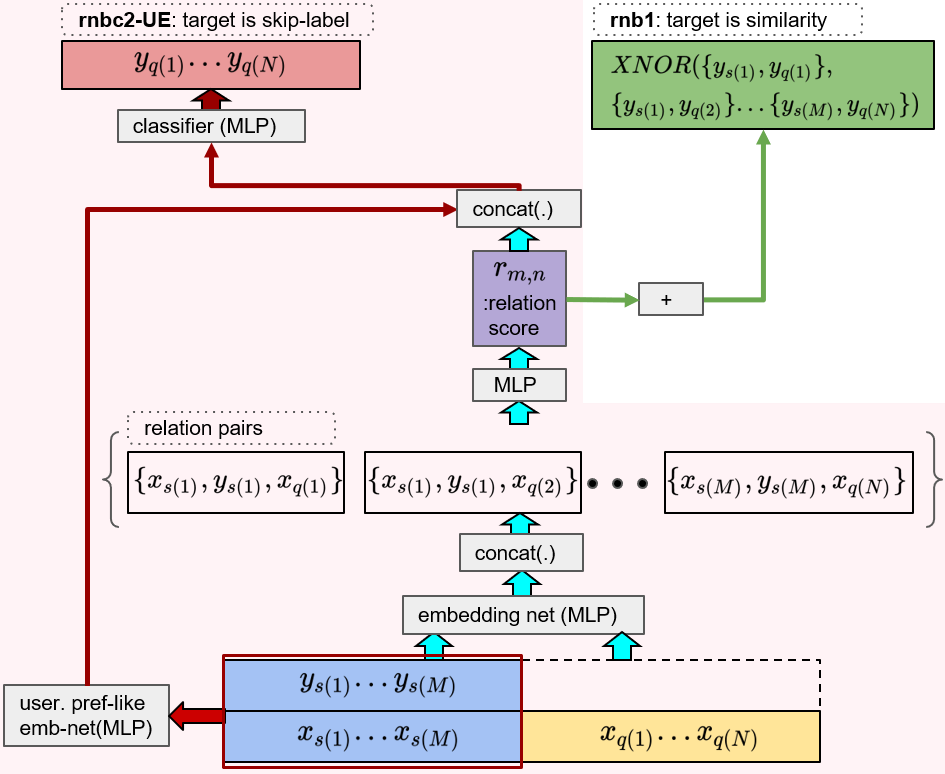}
\caption{``rnb1'' is a relation network-based few-shot metric learner. It can predict a pair-wise similarity (green arrows) by learnt latent metric space: it constructs  all possible relation pairs from the few-shot features and labels. ``rnbc2-UE'' (pink) shares the structure of ``rnb1'', and it can be trained as a few-shot classifier that directly predicts skip-labels.}
\label{rnbc2ue}
\end{figure}

This model aims to learn how to compare a pair of input acoustic features, through a latent metric space, within the context given from the supports. Previously, Sung et al.\cite{sung2018learning} proposed a metric learning for few-shot classification. The relation score $r_{m,n}$ for a pair of support and query inputs, $\{x_{s(m)}, x_{q(n)}\}$ and the label $y_{s(m)}$ can be defined by:
\begin{equation}
    r_{m,n} = \text{RN}(~C(~f_\theta(x_{s(m)}, f_\theta(x_{q(n)}), y_{s(m)}~)~),
\end{equation}
where RN$(.)$ is the relation networks\cite{santoro2017simple}, $f_\theta$ is an MLP for embedding network, and $C(.)$ is a concatenation operator. In the original model\cite{sung2018learning} denoted by \textbf{rnb1}, the sum of the relation score is trained to match the binary target similarity. The target similarity can be computed with {\it XNOR} operation for each relation pair. For example, a pair of items that has same labels will have a target similarity $1$; otherwise $0$. The final model is denoted as \textbf{rnbc2-UE} (Figure \ref{rnbc2ue}) with:
\begin{enumerate}
    \item training the classifier to predict the skip-labels directly, instead of similarity.
    \item trainable parameters to calculate weighted sum of the relation score $r$,
    \item additional embedding layers (the red arrows in Figure 3) to capture the user preference-like.
\end{enumerate}

\begin{figure}
\includegraphics[width=2in]{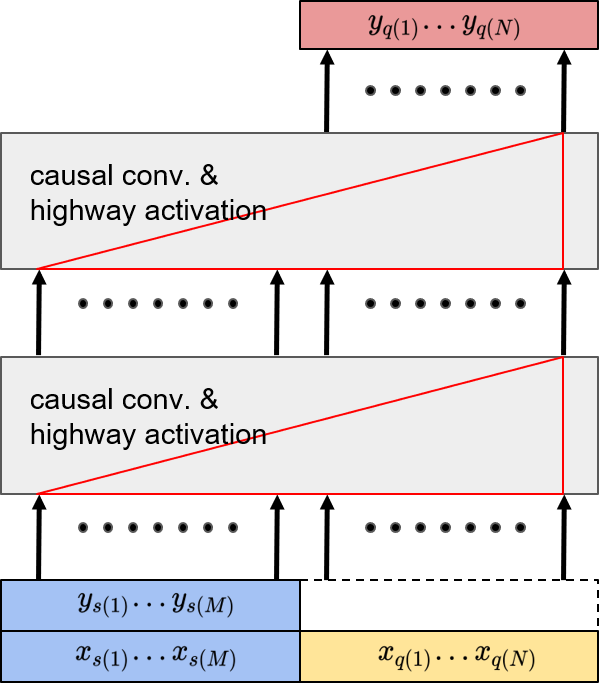}
\caption{``seq1HL'' has 2-stack of causal encoders. A red right triangle represents causal encoder, that is not allowed to observe future inputs.}
\label{seq1hl}
\end{figure}

\subsection{Sequence Learning}
\begin{figure}
\includegraphics[width=2in]{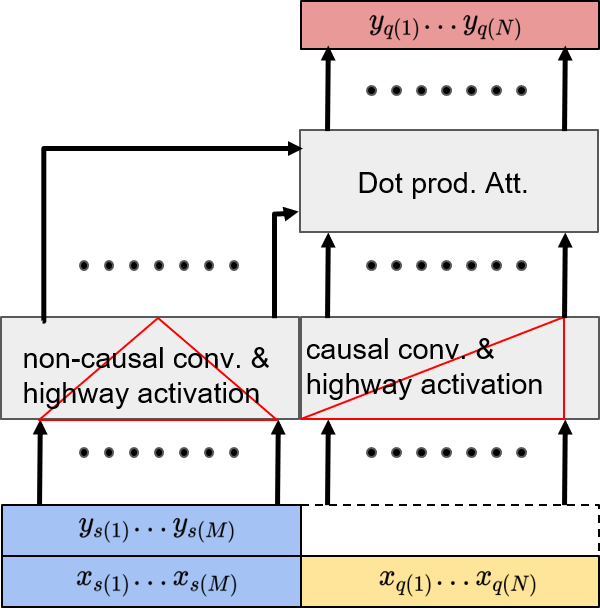}
\caption{``att(seq1eH(S), seq1eH(Q))'' has non-causal encoder for the supports. This allows model to observe future inputs, as represented with a red isosceles triangle.}
\label{att}
\end{figure}
In Figure \ref{seq1hl}, this model consists of dilated convolution layers followed by highway\cite{srivastava2015highway}-activations or GLUs (gated linear units\cite{dauphin2016language}). A similar architecture can be found in the text encoder part of a recent TTS (Text-to-speech) system\cite{dctts}. In practice, we found that non-auto-regressive (non-AR)-models performed consistently better than the AR-models. This was explainable as the noisy outputs of the previous steps degraded the outputs of the next steps cumulatively. The final model, \textbf{seq1HL}, has the following features:
\begin{enumerate}
    \item a non-AR model,
    \item highway-activations with instance norm\cite{vedaldi2016instance}, instead of using GLUs,
    \item $1$-$d$ causal convolution layers with a set of dilation parameters $d = \{1,2,4,8,16\}$ and kernel size $k=2$,
    \item in train, parameters are updated using the loss of $Y_q$, instead of the entire loss of $\{Y_q, Y_q\}$.
\end{enumerate}

We have two variants of the sequence learning model with attention modules. The model in Figure \ref{att} has separate encoders for supports and queries. The support encoder has $1$-stack of non-causal convolution with a set of dilation parameters $d = \{1,3,9\}$ and kernel size $k=3$. The query encoder has $1$-stack of causal convolution with a set of dilation parameters $d = \{1,2,4\}$  and kernel size $k=\{2,2,3\}$. These encoders are followed by a dot product attention operation\cite{vaswani2017attention}. 

\begin{figure}
\includegraphics[width=2in]{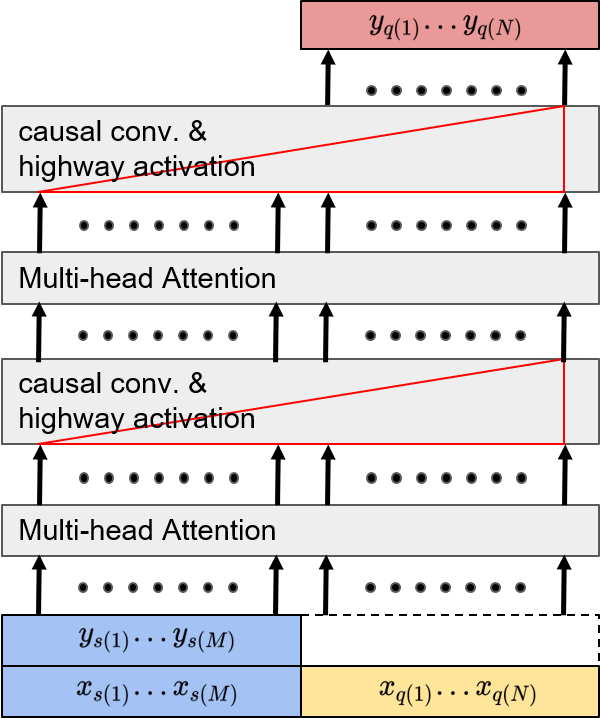}
\caption{SNAIL\cite{snail}-like model. We removed the first embedding layer, and trained it as a non-AR model.}
\label{snailf}
\end{figure}

In contrast with the models mentioned above, SNAIL\cite{snail} (in Figure \ref{snailf}) has attention module at the bottom, and the causal convolution layer follows. For the multi-head attention, we set the number of head to 8.  

\section{Experiments}
\subsection{Pre-processing}
From the {\it Spotify} dataset\cite{brost2019music}, we decoded the categorical text labels in session logs into one-hot vectors. Other integer values from the logs, such as ``number of times the user did a seek forward within track'' were min-max normalized after taking logarithm. We didn't make use of dates. The acoustic  features were standardized to have mean=$0$ with std=$1$.
\subsection{Evaluation Metric}
The primary metric for the challenge was {\it Mean Average Accuracy} (MAA), with the average accuracy defined by
\begin{math}
    AA = \sum_{i=1}^T A(i)L(i) / {T},
\end{math} 
where $T$ is the number of tracks to be predicted for the given session, $A(i)$ is the accuracy at position $i$ of the sequence, and $L(i)$ is the boolean indicator for if the $i$-th prediction was correct.
\subsection{Training}
In all experiments displayed in Table 1, we trained the models using $80$\% of train set. The rest of train set was used for  validation. \textbf{rnb1} and \textbf{rnb2-UE} was trained with MSE loss. All other models were trained with binary cross entropy loss.  We used Adam\cite{kingma2014adam} optimizer with learning rate $10^{-3}$, annealed by 30\% for every 99,965,071 sessions (= 1 epoch). Every training was stopped within 10 epochs, and the training hour varied from 20 to 48. We uniformly applied the batch-size 2,048. For the baseline algorithms that have not been submitted, we display the validation MAA instead. The total size of  trainable parameters for each model can vary. For comparison of model architectures, we maintained the in-/output dimensions of every repeated linear units in metric learning as 256. In sequence learning, we maintained the size of in-/output channels as 256 for every encoder units.
\begin{table}
\small
\begin{threeparttable}
  \caption{Main Results}
  \label{mainresult}
  \begin{tabular}{lccl}
    \toprule
    Model & Category & MAA(ofc) & MAA(val)\\
    \midrule
    rnb1  & M & - & 0.540\\
    rnb2-UE  & M & - & 0.564\\
    rnbc2-UE & M &0.574 & 0.574\\
    \midrule
    seq1eH (1-stack) & S &0.633 & 0.633\\
    seq1HL (2-stack)& S & \textbf{0.637} & \textbf{0.638}\\
    att(seq1eH(S), seq1eH(Q))& S & -  & 0.633\\
    self-att. transformer & S & -  & 0.631\\
    replicated-SNAIL & S & - &0.630\\
  \bottomrule
\end{tabular}
    \begin{tablenotes}
      \footnotesize
      \item \textbf{MAA(ofc)} from official evaluation; \textbf{MAA(val)} from our validation; \textbf{M} and \textbf{S} denote metric and sequence learning, respectively; \textbf{rnb1} was the replication of ``learning to compare''\cite{sung2018learning}; \textbf{rnbc2-UE} and \textbf{seq1HL} were our final model for metric and sequence learning, respectively; 
      
    \end{tablenotes}
\end{threeparttable}
\end{table}
\subsection{Main Results and Discussion}
Note that we only discuss here the results from non-AR setting. The main results are displayed in Table 1. We can compare the metric learning-based algorithms in the first three rows. \textbf{rnb1} was the firstly implemented algorithm. \textbf{rnb2-UE} had two additional embedding layers. It achieved 2.4\%p improvements over \textbf{rnb1}. The final model, \textbf{rnbc2-UE} additionally achieved 1\%p improvements by changing the target label from similarity to skip-labels.

The five rows from the bottom  display the performance of sequence learning-based algorithms. \textbf{seq1eH} and \textbf{seq1HL} shared the same architecture, but differed in the depth of the networks. \textbf{seq1HL} achieved the best result, and it showed 0.5\%p improvement over \textbf{seq1eH}. \textbf{att(seq1eH(S), seq1eH(Q))} showed a comparable performance with \textbf{seq1eH}. The trans-former\cite{vaswani2017attention} and SNAIL\cite{snail} were also attention-based models. However, we could observe that sequence learning-based model without attention unit worked better. 

Overall, the sequence learning-based approaches outperformed the metric learning-based approaches by at least 5.9\%p. The large difference in performance implied that sequence learning was more efficient, and the metric learning-based models were missing crucial information from the sequence data.

\subsection{How helpful would it be if complete information was provided to query sets?}
\begin{table}
\small
  \caption{The effect of complete information provided to query}
  \label{tab:freq}
  \begin{tabular}{ccccl}
    \toprule
    Model&User-logs & Acoustic feat. & Skip-label & MAA(val) \\
    \midrule
    Teacher&use  & use & - & 0.849\\
    seq1HL&-   & use & - & 0.638\\

  \bottomrule
\end{tabular}

\end{table}
So far, the input query set $X_q$ has been defined as acoustic features (see Figure 1). In this experiment, we trained a new model \textbf{Teacher} using both user-logs and acoustic features that were available in dataset. In Table 2, the performance of the \textbf{Teacher} was 21.1\%p higher than our best model \textbf{seq1HL}. This revealed that the user-logs for $X_q$ might contain very useful information for sequential skip prediction. In future work, we will discover how to distill the knowledge.

\section{Conclusions}
In this paper, we have described two different approaches to solve the sequential skip prediction task with few-shot in online music service. The first approach was based on metric learning, which aimed to learn how to compare the music contents represented by a set of acoustic features and user interaction logs. The second approach was based on sequence learning, which has been widely used for capturing temporal information or learning how to refer past experience. In experiments, our models were evaluated in WSDM Cup 2019, using the real dataset provided by {\it Spotify}. The main results revealed that the sequence learning approach worked consistently better than metric learning. In the additional experiment, we verified that giving a complete information to the query set could improve the prediction accuracy. In future work, we will discover how to generate or distill these knowledge by the model itself. 

\begin{acks}
\small
This work was supported by Kakao and Kakao Brain corporations, and by  National Research Foundation (NRF2017R1E1A1A01076284).
\end{acks}

\bibliographystyle{ACM-Reference-Format}
\bibliography{ref_skip}
\end{document}